\documentclass{aastex61}

\begin{document}

\bibliographystyle{apj}

\title{Constraining Hot Jupiter Atmospheric Structure and Dynamics through Doppler Shifted Emission Spectra}

\author{Jisheng Zhang}

\affil{Department of Physics, Grinnell College, 1116 8th Ave. Grinnell, IA 50112, USA}

\email{zhangjis17@grinnell.edu}

\author{Eliza M.-R. Kempton}

\affil{Department of Physics, Grinnell College, 1116 8th Ave. Grinnell, IA 50112, USA}

\email{kemptone@grinnell.edu}
  
\author{Emily Rauscher}

\affil{Department of Astronomy, University of Michigan, 311 West Hall, 1085 South University, Ann Arbor, MI 48109, USA}

\begin{abstract}

We present a coupled 3-D atmospheric dynamics and radiative transfer model to predict the disk-integrated thermal emission spectra of transiting exoplanets in edge-on orbits. We calculate spectra at high resolution to examine the extent to which high-resolution emission spectra are influenced by 3-D atmospheric dynamics and planetary rotation, and to determine whether and how we can constrain thermal structures and atmospheric dynamics through high-resolution spectroscopy. This study represents the first time that the line-of-sight geometry and resulting Doppler shifts from winds and rotation have been treated self-consistently in an emission spectrum radiative transfer model, which allow us to assess the impact of the velocity field on thermal emission spectra. We apply our model to predict emission spectra as a function of orbital phase for three hot Jupiters, HD 209458b, WASP-43b and HD 189733b. We find net Doppler shifts in modeled spectra due to a combination of winds and rotation at a level of 1-3 km/s. These Doppler signatures vary in a quasi-sinusoidal pattern over the course of the planets' orbits as the hot spots approach and recede from the observer's viewpoint. We predict that WASP-43b produces the largest Doppler shift due to its fast rotation rate. We find that the net Doppler shift in an exoplanet's disk-integrated thermal emission spectrum results from a complex combination of winds, rotation, and thermal structure. However, we offer a simple method that estimates the magnitude of equatorial wind speeds in hot Jupiters through measurements of net Doppler shifts and lower resolution thermal phase curves.

\end{abstract}


\section{Introduction \label{intro}}

Close in gas-giants, i.e.~hot Jupiters, are ideal candidates for exoplanet atmospheric research. They constitute the largest and most irradiated type of exoplanet, resulting in high planet-to-star flux ratios. In addition, they have very short orbital periods (1-10 days), which means they have a high probability of transiting their host stars, thus making them the targets of many observational efforts. With semimajor axes of a small fraction of an A.U., hot Jupiters reside in extreme and exotic environments. Typically, these gas giants have surface temperatures exceeding 1,000 K.  They are believed to be tidally locked into synchronous rotation states, leaving one side in permanent darkness and the other in a permanent day. These features make hot Jupiters excellent laboratories for studying atmospheric chemistry, structure, and dynamics in a regime that is not accessible to Solar System studies.

The atmospheric and temperature structures of hot Jupiters are inherently three-dimensional. However, with these objects being tens of parsecs away and orbiting luminous host stars, it is impossible to spatially resolve the planets from their host stars with current technology. Therefore, exoplanet astronomers must rely on spectroscopic techniques that disentangle the contribution of the planet from that of the star. Typically, astronomers have used low- to mid-resolution spectroscopy in order to achieve high signal-to-noise (S/N) ratios. For transiting exoplanets, spectra obtained during transit and secondary eclipse, along with a smaller number of full orbital phase curves, have provided constraints on chemical compositions of the atmospheres and temperature structures \citep[e.g.][]{cha02, ric06, cha08, knu09, ste14, kre14b}.

More recently, high-resolution spectroscopy (HRS) at $R \sim$ 10$^4$ - 10$^5$ from ground-based observing platforms has emerged as a complementary observing technique to constrain key exoplanet properties.  In addition to being a useful tool for determining the chemical composition and thermal structure of exoplanet atmospheres, HRS observations can directly probe effects that are only discernible at high resolution, such as atmospheric winds and a planet's rotation.  The HRS technique benefits from the exoplanet being \emph{spectrally} resolvable from both its host star and telluric absorption, due to Doppler shifts from the high orbital velocity of the planet.  While high-resolution spectroscopy has the potential to be a powerful tool for studying exoplanet atmospheres, it requires many more photons than low-resolution spectroscopy to achieve high S/N. Due to this limitation, the tool has so far been limited to very bright stars hosting close-in giant planets, and cross-correlation techniques have typically been required to interpret low S/N spectra. The former issue will likely be resolved with the installation of the next generation of 30-m class telescopes.

The original attempt at HRS characterization of the atmosphere of a hot Jupiter dates back to 1999, when \citet{cha99} used the technique to search for the reflected light signal of $\tau$ Bo\"otis b at optical wavelengths. The team obtained a null result, which is now attributed to the very low albedos of hot Jupiters \citep[e.g.][]{row08}. More recently, however, the HRS technique has been applied at IR wavelengths to great success.  \citet{sne10} obtained the first high-resolution transmission spectrum of HD 209458b at 2.3 $\mu$m, and reported the direct detection of the orbital motion of the planet along with a tentative detection of an additional blueshift that could be attributed to high-altitude winds resulting from heat redistribution from the planet's dayside to its night side \citep{mil12,sho13}. This result is further upheld by similar detections of $\sim$2 km/s blueshifts in the transmission spectrum of HD 189733b \citep{lou15, bro16}.  Subsequent dayside thermal emission measurements with HRS for the transiting planets HD 209458b \citep{sch15, bro17} and HD 189733b \citep{rod13,dek13,bir13} have allowed for the detection of molecules and constraints to be placed on the planets' vertical thermal structures.  Observations of directly emitted light from exoplanets at high resolution additionally enable the characterization of non-transiting exoplanets, because the premise of spectrally resolving the planet from the host star still holds up regardless of the orbital geometry, whereas the low resolution techniques of transmission and secondary eclipse spectroscopy only apply to transiting planets. HRS spectroscopy of $\tau$ Bo\"o b \citep{bro12,rod12,loc14}, 51 Peg b \citep{bro13}, HD 179949b \citep{bro14}, and HD 88133b \citep{pis16} have revealed molecular species and have provided measurements of the true masses of these planets by breaking the $\sin i$ degeneracy inherent to the radial velocity detection technique.  Finally, the HRS technique has been applied to the directly imaged planet, $\beta$ Pictoris b to measure its rotation rate \citep{sne14}.

The interpretation of exoplanet spectra relies on comparison to radiative transfer models (e.g. \citet{bur97}; \citet{sea98}; \citet{bar05}; \citet{for08}). These models have typically been one-dimensional (1-D) meaning the atmosphere is assumed to have the same composition and temperature structure along any radial ray. While the 1-D models have success in matching observations and constraining the chemistry and thermal structures of exoplanets, certain biases could result from standard 1-D assumptions.  For transmission spectra, \citet{bur10} and \citet{for10} generally find good agreement between 1-D and 3-D models, but the calculated spectra diverge significantly near important chemical abundance boundaries. As 1-D models use an averaged T-P profile, the vertical temperature structure uniquely determines the global atmospheric chemistry, under assumptions of thermochemical equilibrium.  However, 3-D models can have strong longitudinal and latitudinal compositional gradients based on the local atmospheric temperature.  For example, in the case of HD 209458b, 1-D models lack TiO absorption features due to a planet-averaged lower temperature, whereas \citet{for10} find gas-phase TiO in abundance on the dayside (hotter region) and a reduced abundance on the planet's limb (cooler region) using a 3-D model.  In thermal emission, \citet{fen16} investigated the bias from 1-D assumptions within a Bayesian atmospheric retrieval framework and showed that chemical abundances could be constrained to incorrect values.  In general, 1-D models lack information on detailed temperature structures and atmospheric dynamics, whereas 3-D models can treat these effects in a more self-consistent manner. 

%\textbf{As 1-D models use an averaged T-P profile, temperature stays either cool enough that certain molecules are condensed out, or hot enough that they are in gas phase. However, 3-D models use a whole range of individual T-P profiles, some of which are hot enough to have certain molecules in gas phase while some are not. Therefore, a 1-D model that does not have spectral features of certain molecules could still correspond to a self-consistent 3-D model that does show these features, because a portion of the atmosphere has these molecules in gas phase. In the case of HD 209458b, 1-D model lacks TiO absorption features due to an averaged lower temperature, whereas \citet{for10} find TiO in abundance on the dayside (hotter region) and a reduced abundance on the planet's limb (cooler region) in the 3-D model.}

3-D general circulation models (GCMs) of the atmospheric dynamics of hot Jupiters have been presented by various authors to investigate detailed thermal and atmospheric structures (e.g. \citet{sho09}; \citet{dob10}; \citet{rau10}; \citet{hen11a}; \citet{thr11}; \citet{may14}). Even though these models differ from each other regarding assumptions such as boundary conditions and the level of complexity of the underlying physics, they mostly make the same primary predictions for the circulation pattern that develops on highly-irradiated tidally-locked planets.  The three key predictions are (1) an equatorial jet at a pressure of $\sim$1 bar moving in the direction of the planet's rotation, (2) the planet's hottest location (hot spot) being shifted eastward from the substellar point due to the aforementioned jet, and (3) \sout{high altitude} day-to-night winds at low pressures. These general trends have been bourne out in a limited number of phase curve and transmission spectrum observations of hot Jupiters \citep[e.g.][]{knu07, sne10}.  Additional data and the development of new observational diagnostics will be required to further affirm the universality of the GCM predictions.  

Motivated by recent observations of high-resolution emission spectra and the need to understand 3-D atmospheric winds and temperature structures, we couple an existing atmospheric dynamics model \citep{rau12} together with a radiative transfer solver to self-consistently incorporate 3-D temperature, pressure, and wind information into the radiative transfer calculation. The final result is to produce thermal emission spectra that are consistent with the underlying 3-D atmospheric structure. This study represents the first time that a full 3-D atmospheric dynamics model has been used as the basis to produce thermal emission spectra at high resolution, whereas previous efforts of this type have only modeled low-resolution or band-averaged spectra \citep{for06, bur10,fen16}.  This is also the first time that the geometry of the line-of-sight radiative transfer through the 3-D atmosphere has been self-consistently treated. Our modeling at high spectral resolution allows for Doppler effects from the planet's rotation and winds to be accounted for, as these effects are inconsequential at low resolution.  

Our goal is to understand the extent to which line-of-sight motions affect HRS observations of hot Jupiter atmospheres and whether information about wind and rotational velocities can be recovered.  We also investigate new diagnostics of 3-D atmospheric temperature structure that will be complementary to low-resolution phase curve observations.  We specifically model three benchmark hot Jupiters, HD 209458b, HD 189733b, and WASP-43b which span a wide range of surface gravity, temperature, and rotational speeds.  Our paper is organized as follows: We describe our 3-D dynamics model and radiative transfer treatment in Section~\ref{methods}. We present our thermal emission spectrum results in Section~\ref{results}. Finally, we provide a summary and conclusions in Section~\ref{conclusion}.    

\section{Model Description \label{methods}}

\subsection{3-D dynamics models}

In order to predict the Doppler shifts in the planetary spectrum that result from motions in the planet's atmosphere, we must self-consistently include both the rotation of the planet and the atmospheric circulation wind pattern.  The planet's winds and rotation are intrinsically linked, in that the Coriolis force is a primary balancing force in the atmospheric circulation; a more quickly rotating planet will develop a different pattern of winds than a more slowly rotating one.  In addition, the detailed thermal structure of the atmosphere is shaped by the relative balances between radiative heating and heat transport through advection.  All of these physical processes are included in the three-dimensional atmospheric circulation model that we use to predict the temperature and wind structure of these planets.  For full details of the model, see \citet{rau10,rau12}.  In short, it is a numerical code that solves the ``primitive equations of meteorology" (the fluid equations, subject to simplifying assumptions relevant to an atmosphere on a rotating sphere) coupled with radiative heating/cooling via ``double-gray" two-stream radiative transfer (separating the incoming stellar flux from thermal re-emission, using a single absorption coefficient for each).

We model each planet over a pressure range from 10 microbar to 100 bar (with 45 levels evenly spaced in log pressure), a horizontal spectral resolution of T31 (corresponding to a spatial resolution of $\sim$3.75\degr), and using the parameters listed in Table 1.  (We assume each planet has been tidally locked into synchronous rotation, such that its orbital and rotation periods are equal.)  We chose values for the optical and infrared absorption coefficients that produce analytic temperature-pressure profiles \citep{gui10} that roughly match the results from more detailed one-dimensional radiative transfer models. Each simulation was initialized with a temperature structure that was horizontally uniform and varied vertically according to the analytic double-gray solution.  The atmospheres began with zero winds (in the frame rotating with the planet) and all simulations ran for 2000 orbits, at which point all but the deepest (unobservable) regions of the atmosphere had accelerated to a statistically steady state.  

\begin{deluxetable}{lccc}
\tablecaption{Planet parameters}
\tablehead{
  \colhead{}                                            &  \colhead{HD 209458b}            & \colhead{WASP-43b}                   & \colhead{HD 189733b}
}
\startdata
Radius of the planet, $R_p$                   & $1.0 \times 10^8$ m                   & $7.4 \times 10^7$ m               & $8.7 \times 10^7$ m \\
Gravitational acceleration, $g$              & 8.0 m s$^{-2}$                              & 47.0 m s$^{-2}$                         & 19.5 m s$^{-2}$ \\
Rotation rate, $\Omega$                       & $2.1 \times 10^{-5}$ s$^{-1}$    & $8.9 \times 10^{-5}$ s$^{-1}$    & $3.3 \times 10^{-5}$ s$^{-1}$ \\
Incident flux at substellar point, $F_0$ & $1.06 \times 10^6$ W m$^{-2}$ & $9.78 \times 10^5$ W m$^{-2}$  & $4.74 \times 10^5$ W m$^{-2}$   \\
\ \ \ Corresponding temperature, $T_{\mathrm{irr}}$      & 2080 K                 & 2040 K                                        & 1700 K\\
Optical absorption coefficient, $\kappa_{\mathrm{vis}}$ & \multicolumn{3}{c}{$4 \times 10^{-3}$ cm$^2$ g$^{-1}$} \\
\ \ \ Optical photosphere ($\tau_{\mathrm{vis}}=2/3$)   & 130 mbar             & 780 mbar                                    & 330 mbar \\
Infrared absorption coefficient, $\kappa_{\mathrm{IR},0}$ & \multicolumn{3}{c}{$1 \times 10^{-2}$  cm$^2$ g$^{-1}$} \\
\ \ \ Infrared photosphere ($\tau_{\mathrm{IR}}=2/3$)   & 53 mbar               & 310 mbar                                    & 130 mbar \\
\enddata
\end{deluxetable}

\subsection{Emission Spectrum Radiative Transfer}
  
\begin{figure*}
\gridline{\fig{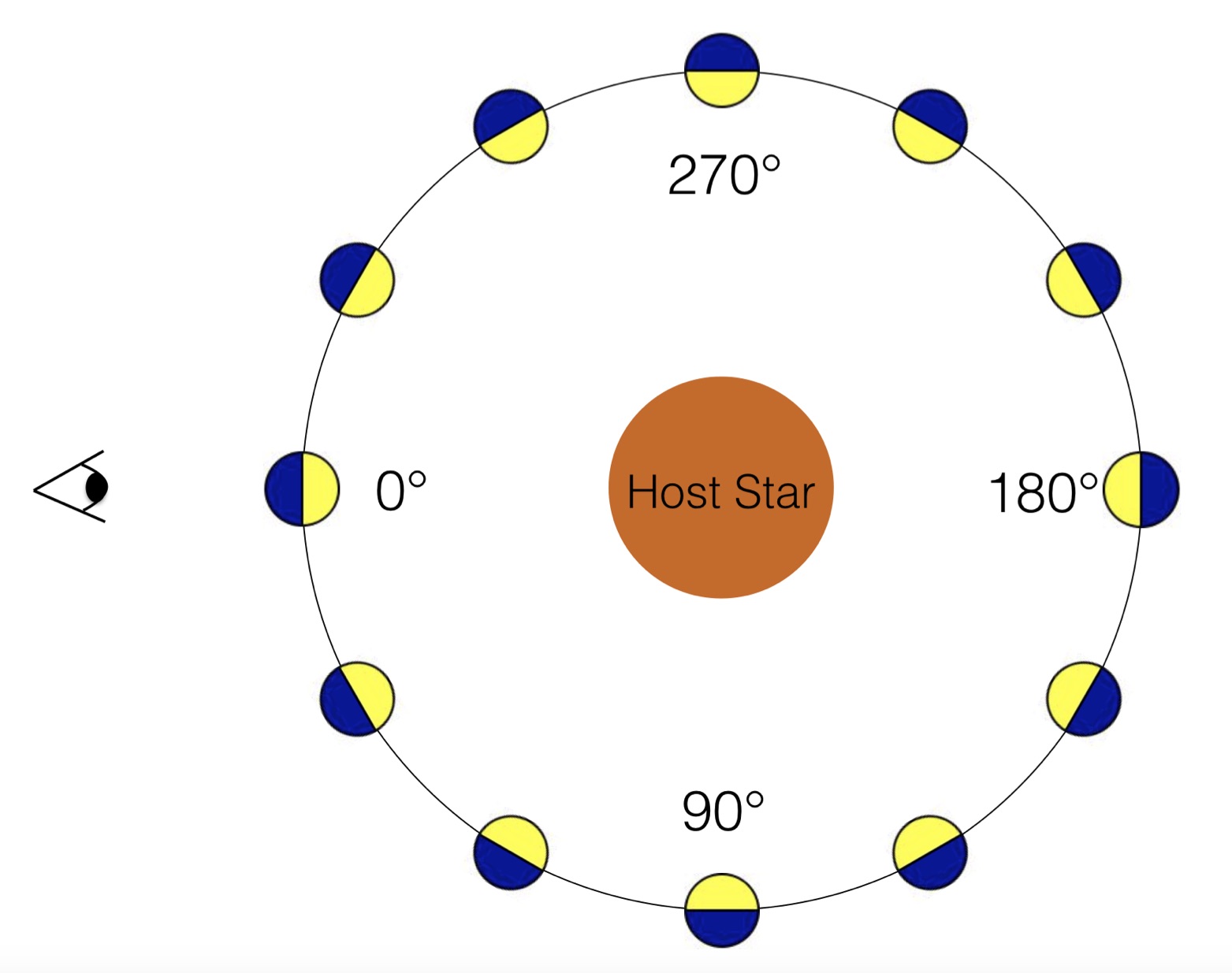}{0.3\textwidth}{(a)}
          \fig{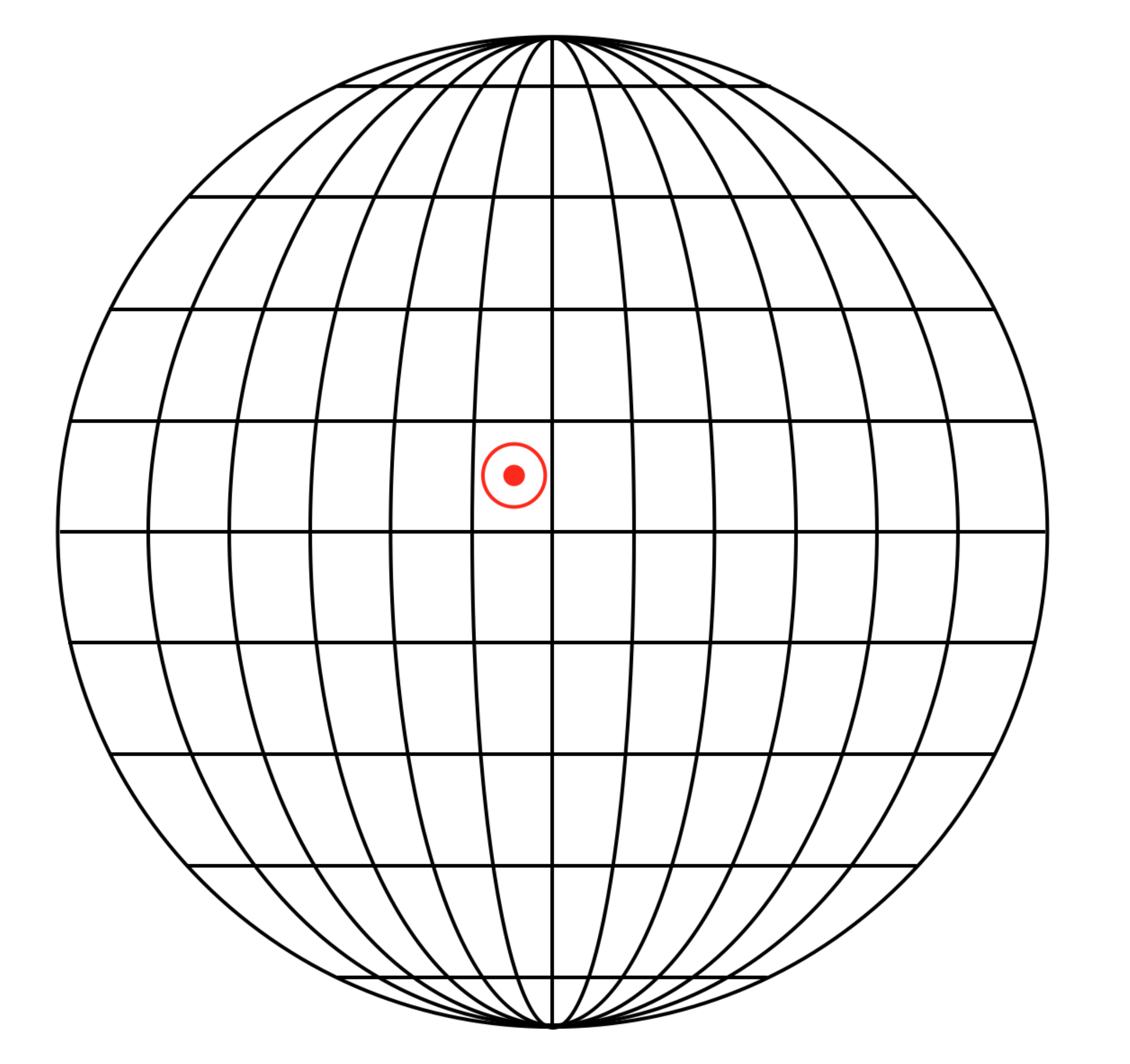}{0.25\textwidth}{(b)}
          \fig{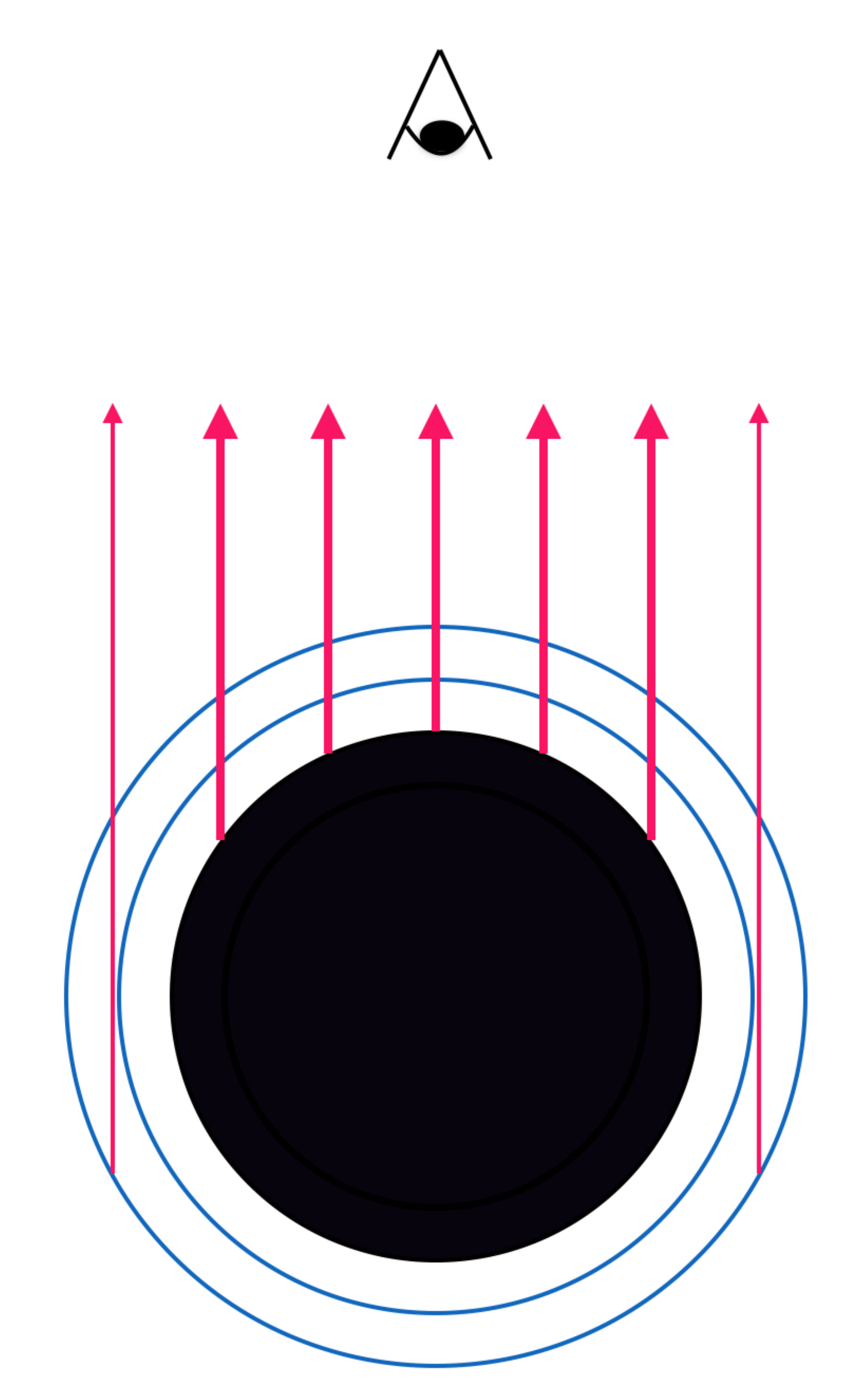}{0.235\textwidth}{(c)}
          }
\caption{The geometry of the 3-D thermal emission problem. Panel (a) shows the orbital motion of the planet viewed at a 90$^{\circ}$ angle to the orbital plane. The planet is depicted at 12 different phase angles from $\varphi = 0^{\circ}$ to $360^{\circ}$, in $30^{\circ}$ increments. Yellow indicates the planet's permanent dayside, whereas the blue denotes the (cooler) night-side. Panel (b) shows a face-on view of the planet's latitude-longitude grid as well as the geometry of a single light ray emanating from a grid cell along the observer's sightline. Panel (c) shows a top-down view of the line-of-sight light rays. The black circle represents the planetary interior, whereas the blue rings indicate each layer of the atmosphere. Red arrows shows the pathway of light rays, with thinner arrows originating from the limb and thicker ones from the bottom layer of the atmosphere. Note that this panel is not drawn to scale and that we neglect the minor contribution of rays originating at the limb in our calculations.   \label{fig:2017-03-12_7_09_43}}
\end{figure*}

   From the GCM results, the emission spectrum is obtained by summing the intensity of all line-of-sight light rays emergent from the visible side of the hot Jupiter. We calculate the intensity of a single light ray passing through the planetary atmosphere by solving the radiative transfer equation for pure thermal emission according to 
\begin{equation}
I(\lambda)=B_0e^{-\tau_0}+\int_{0}^{\tau_0}e^{-\tau}B d\tau,
\label{eqn1}
\end{equation}
where $B_0$ is the Planck function evaluated at the local temperature of the deepest layer of the atmosphere, $B$ is the thermal emission (Planckian) source function evaluated at the local gas temperature, $\tau$ is the slant optical depth (described below), and $\tau_0$ is the slant optical depth at the base of the atmosphere.  Note that all quantities expressed in Equation~\ref{eqn1} are implicitly wavelength dependent.  The first term of Equation~\ref{eqn1} accounts for the initial intensity diminished by absorption, and the second term accounts for the intensity of integrated thermal emission along the pathway, attenuated by absorption. We ignore effects of multiple scattering and treat Rayleigh scattering as an absorption cross section, which effectively assumes that all scattered light is removed from the beam intersecting the observer's sight line. To calculate the integrated flux, we divide the visible hemisphere into 2,304 individual patches set by the latitude-longitude grid of the GCM and propagate a line-of-sight light ray through each one (see Figure~\ref{fig:2017-03-12_7_09_43}). The emergent intensity of each ray is calculated according to Equation~\ref{eqn1}, and then integrated over the corresponding solid angle subtended by each grid cell to obtain the total flux. 

The slant optical depth, $\tau$, is evaluated according to 
\begin{equation}
\tau=\int{\kappa}ds,
\label{eqn2}
\end{equation}
where d$s$ is the differential path length of each layer of the atmosphere along the observer's line of sight, and $\kappa$ is the opacity evaluated at the local temperature and pressure, which in turn depend on the local latitude, longitude and altitude. We account for absorption and Rayleigh scattering opacities for gas phase CH$_4$, CO, CO$_2$, H$_2$O, and NH$_3$, as well as collision-induced H$_2$ opacities \citep{fre08}.  Chemical abundances are set by an atmosphere of solar composition gas, in a state of local thermochemical equilibrium.  

For calculations at high spectral resolution (R $>$ 10$^5$), the opacity is Doppler shifted by the local line-of-sight wind velocity and rotational velocity of the planet.  In our model, we exclude the contribution from the orbital motion, as it is typically a known quantity that can easily be subtracted off from the total Doppler shift. (We discuss this assumption in more depth in Section~\ref{results341})  The remaining line-of-sight velocity is given by 
\begin{equation}
v_{LOS}=-[u\sin({\phi + \varphi})\cos{\theta}+v\cos({\phi+\varphi})\sin{\theta}-w\cos({\phi + \varphi})\cos{\theta}+(R_p+z)\Omega\sin({\phi + \varphi})\cos{\theta}].
\label{eqn:vlos}
\end{equation}
Here, $\theta$ and $\phi$ are the latitude and longitude ($\theta = \phi = 0$ defines the substellar point), and $\varphi$ is the orbital phase angle, with $\varphi = 0$ corresponding to the time when the full night side of the planet is in view. The first three terms in the equation give the contribution to the line-of-sight velocity from the atmospheric motion, where $u$, $v$, and $w$ are the zonal (east-west), meridional (north-south), and vertical component of the line-of-sight wind speed, respectively, (where east, north, and up define the positive direction). The fourth term provides the contribution from the rotation of the planet, where $R_p$ is the radius of the planet at the base of the atmosphere, $z$ is the altitude of each layer above the base of the atmosphere, and $\Omega$ is the planet's rotational speed in rad/s, assuming synchronous rotation. Our model assumes a circular orbit for the hot Jupiter with an orbital inclination of exactly 90$^{\circ}$. For our high resolution calculations, the opacity, $\kappa$, in Equation~\ref{eqn2} is then evaluated at the Doppler-shifted wavelength 
\begin{equation}
\lambda=\lambda_0(1-\frac{v_{LOS}}{c})
\end{equation}
where $\lambda_0$ is the unshifted wavelength, thus incorporating atmospheric dynamics self-consistently into our simulated thermal emission spectra.

This paper marks the first time that the line-of-sight radiative transfer has been modeled in a geometrically consistent manner.  Previous studies of thermal emission spectra based on 3-D GCMs \citep{for06,bur10} performed radiative transfer calculations along radial rays and used versions of the plane parallel approximation to model the limb-darkening effect. Our current work necessitates that we explicitly pass light rays through the atmosphere along the observer's sightline, so that the Doppler shifts associated with line-of-sight atmospheric motions are correctly accounted for.  In addition to performing the line-of-sight radiative transfer calculation over the disk of the planet as defined by the GCM grid, we have also run tests in which we calculate the added contribution from rays originating at the planet's limb, i.e. those rays which pass entirely through the atmosphere from back to front and do not encounter the deepest layer of the atmosphere.  The geometry of the limb calculation is the same as the one described for transmission spectra in \citet{mil12}, except that here the contribution from thermal emission (rather than absorption of stellar light) is considered.  The solid angle subtended by the limb is small when compared to the disk of the planet ($\sim 2 H/R_p$, where $H$ is the pressure scale height), but motions associated with winds and rotation in this region are well-aligned with the observer's sightline.  Our numerical calculations reveal that the limb only contributes up to 0.5$\%$ of the total flux of the planet and does not significantly alter the results described in the following sections.  We therefore neglect the contribution from the limb in our subsequent modeling.

We produce model emission spectra at both low and high spectral resolution. Low-resolution spectra ($R = 1,000$; with no Doppler shifts) are modeled over the visible to mid-IR from 0.5 to 12 $\mu m$. High-resolution spectra ($R = 10^6$; including Doppler shifts) are calculated from 2308 to 2314 nm, to specifically sample a representative portion of the wavelength range over which thermal emission spectra have been obtained with the CRIRES spectrograph.  This narrow wavelength range includes a prominent CO absorption band along with H$_2$O and CH$_4$ spectral features. We expect the wavelength range of our HRS models to have minimal impact on our results, with the following caveats.  Wavelength ranges that primarily produce stronger (or weaker) absorption features will preferentially probe higher (or lower) altitudes in the atmosphere.  Similarly, absorption feautures associated with certain chemical species that are non-uniformly distributed around the planet will preferentially probe the locations where those species are the most abundant.  We choose to focus on the 2.3 $\mu$m CO bandhead because the vibration-rotation spectrum of CO is well-characterized, and prior HRS observations have been obtained at these wavelengths.

%\textbf{The challenge for observing other wavelength ranges that probe molecules, such as CH$_4$ and CO$_2$, is that the spectra of other molecules are much more complex and less well-understood. Since the technique relies on a template, against which we cross-correlate observations, not having the spectra exactly correct will limit the ability to detect molecules in the atmosphere and constrain atmospheric structures.}

Our modeling implicitly assumes that the stellar spectrum remains stable, and we do not account for sources of stellar variability. This should be a sound assumption from an observational standpoint for the following reasons.  Because stellar rotation periods are typically much longer than the orbital periods of hot Jupiters, and starspots do not evolve substantially over the timescale of making an observation, stellar variability is not expected to have a strong effect on exoplanetary emission spectra. Furthermore, at high spectral resolution, the large Doppler shift of the planetary spectrum relative to the stellar spectrum should allow for full removal of the stellar lines.  Care should be taken, however, in stitching together phase curves from observations taken over timescales comparable to the stellar rotation period or in situations where the planet and star have overlapping absorption spectral lines.

\section{Results \label{results}}

\subsection{GCM Results \label{gcm_results}}

\begin{figure*}
\gridline{\fig{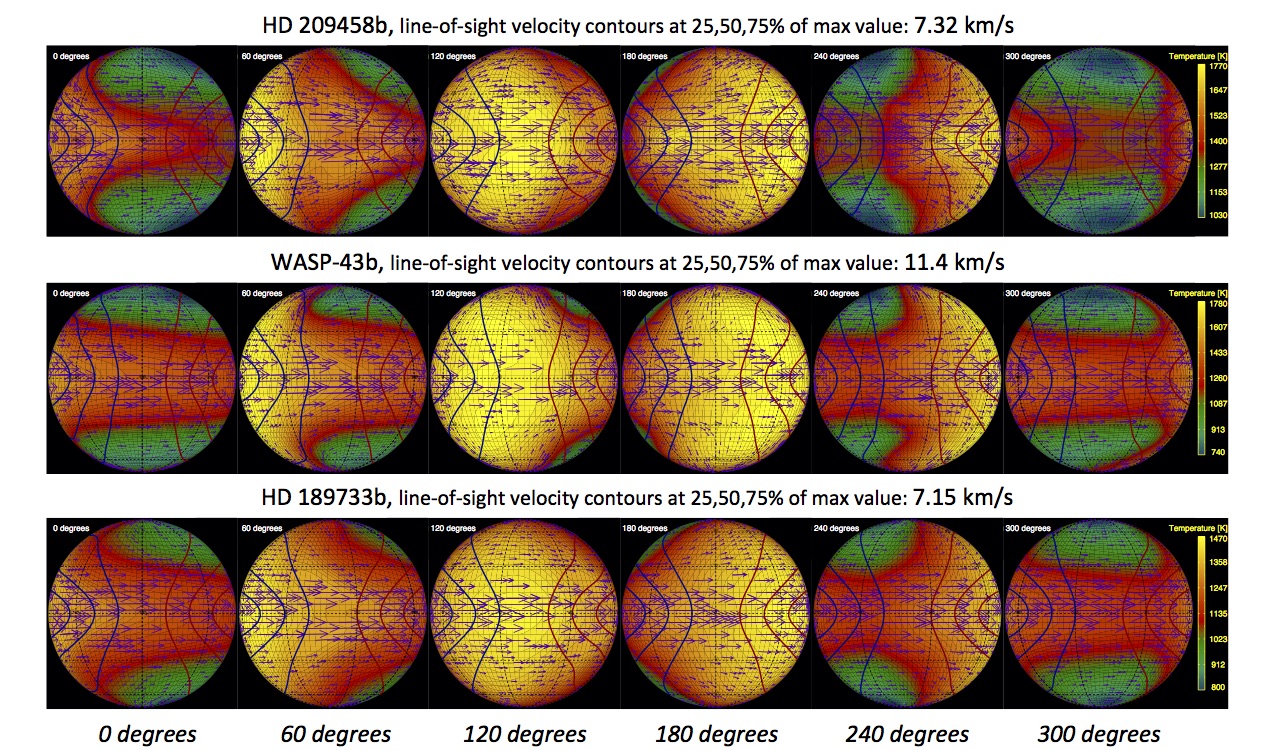}{0.75\textwidth}{}
          }
\gridline{\fig{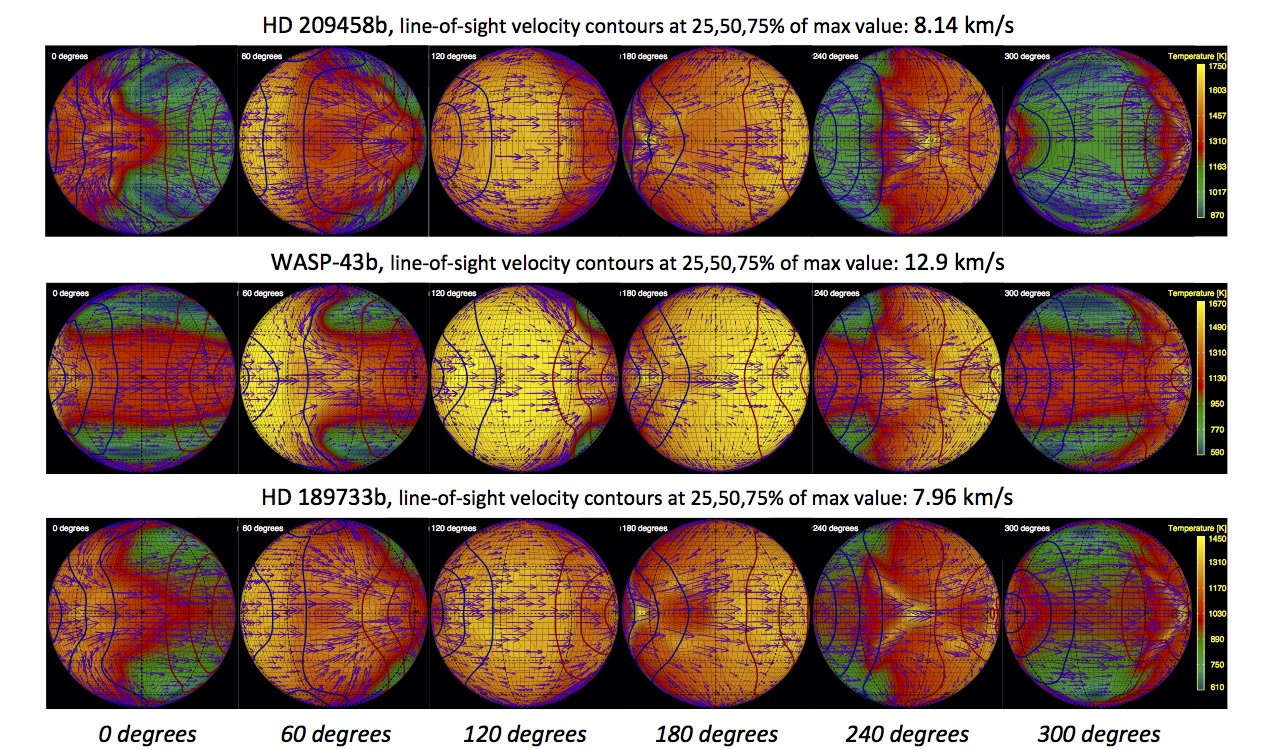}{0.75\textwidth}{}
          }
\caption{Snapshots of three planets at IR photosphere pressure levels (top; 53 mbar, 310 mbar, and 130 mbar for HD 209458b, WASP-43b, and HD 189733b, respectively) and lower pressure levels corresponding to the strongest absorption lines in the post-processed emission spectra (bottom; 0.14 mbar, 0.85 mbar, and 0.35 mbar, respectively) at 6 orbital phase angles, with two-dimensional wind vectors and line-of-sight velocity contours including both winds and rotation (red for redshift and blue for blueshift) overplotted. In each panel, temperature is indicated by the color scale.
\label{fig:snapshot}}
\end{figure*}

Each of the planets we model shows a standard hot Jupiter circulation pattern, as can be seen in the maps of temperature and wind structures shown in Figure \ref{fig:snapshot}.  Each planet has an eastward jet along the equator and eastward advection of the hottest region of the atmosphere away from the substellar point.  There is also a component of day-to-night flow over the poles, which is more pronounced at lower pressure levels (higher in the atmosphere).  The temperature structures also show more complex features at lower pressures, the result of stronger stellar heating and faster winds at those levels.  In Figure \ref{fig:snapshot} we also overplot contours of the line-of-sight velocity toward the observer, a combination of each planet's global rotation and the local wind speeds.  The equatorial rotational velocity of each planet is: 2.1 km/s, 6.6 km/s, and 2.8 km/s for HD 209458b, WASP-43b, and HD 189733b, respectively.  Within the rotating reference frame of each respective planet, the peak wind speed at the pressure levels associated with the CO line core are: 6.1 km/s, 6.3 km/s, and 5.1 km/s.  While all of the planets have similar circulation patterns and wind speeds, WASP-43b has the strongest line-of-sight velocities because of its much faster rotation.  Little-to-no variability is observed in each model; the effect on line-of-sight velocities, for example, is less than 1\%.

\begin{figure*}
\gridline{\fig{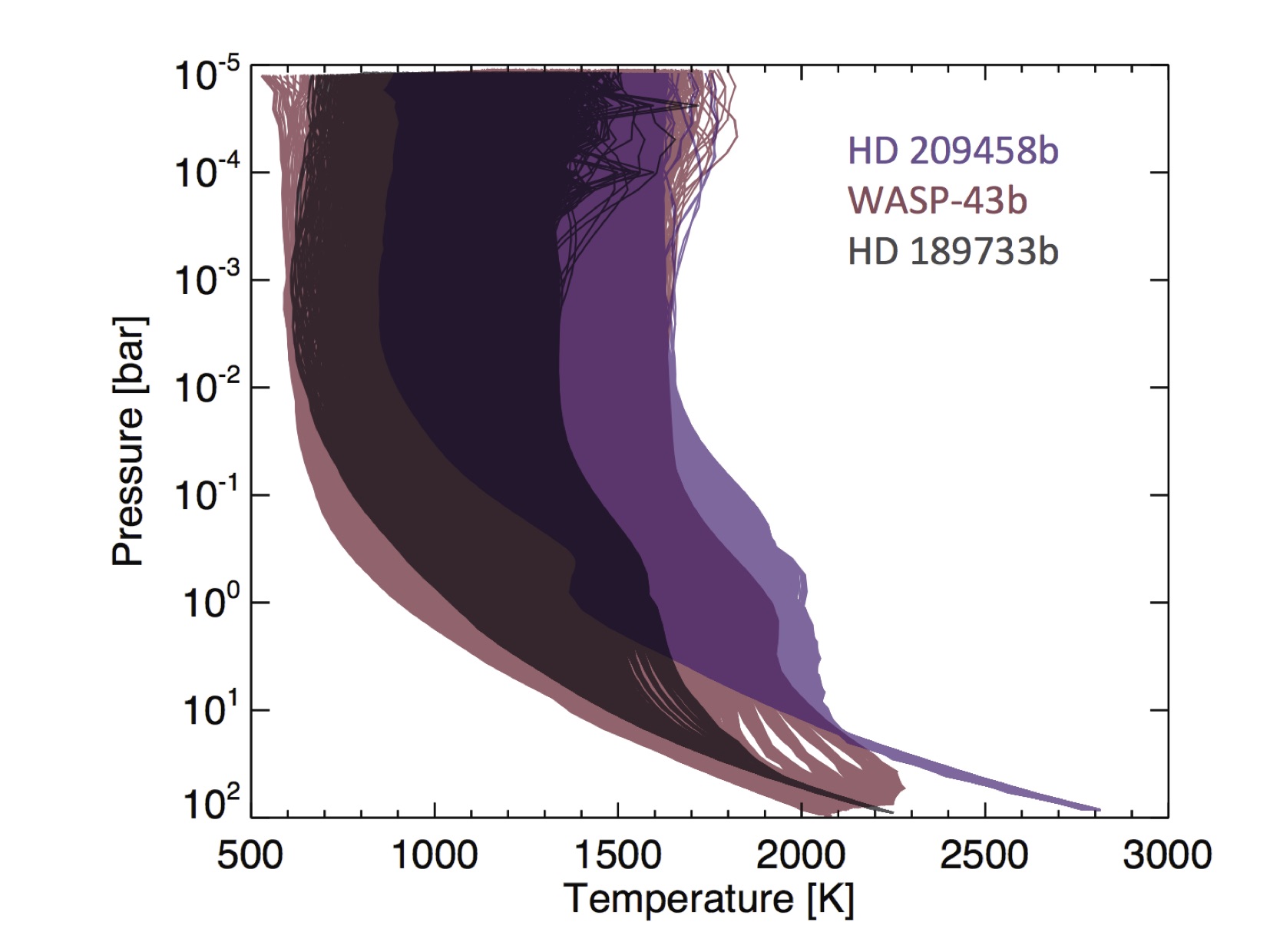}{0.75\textwidth}{}
         }
\gridline{\fig{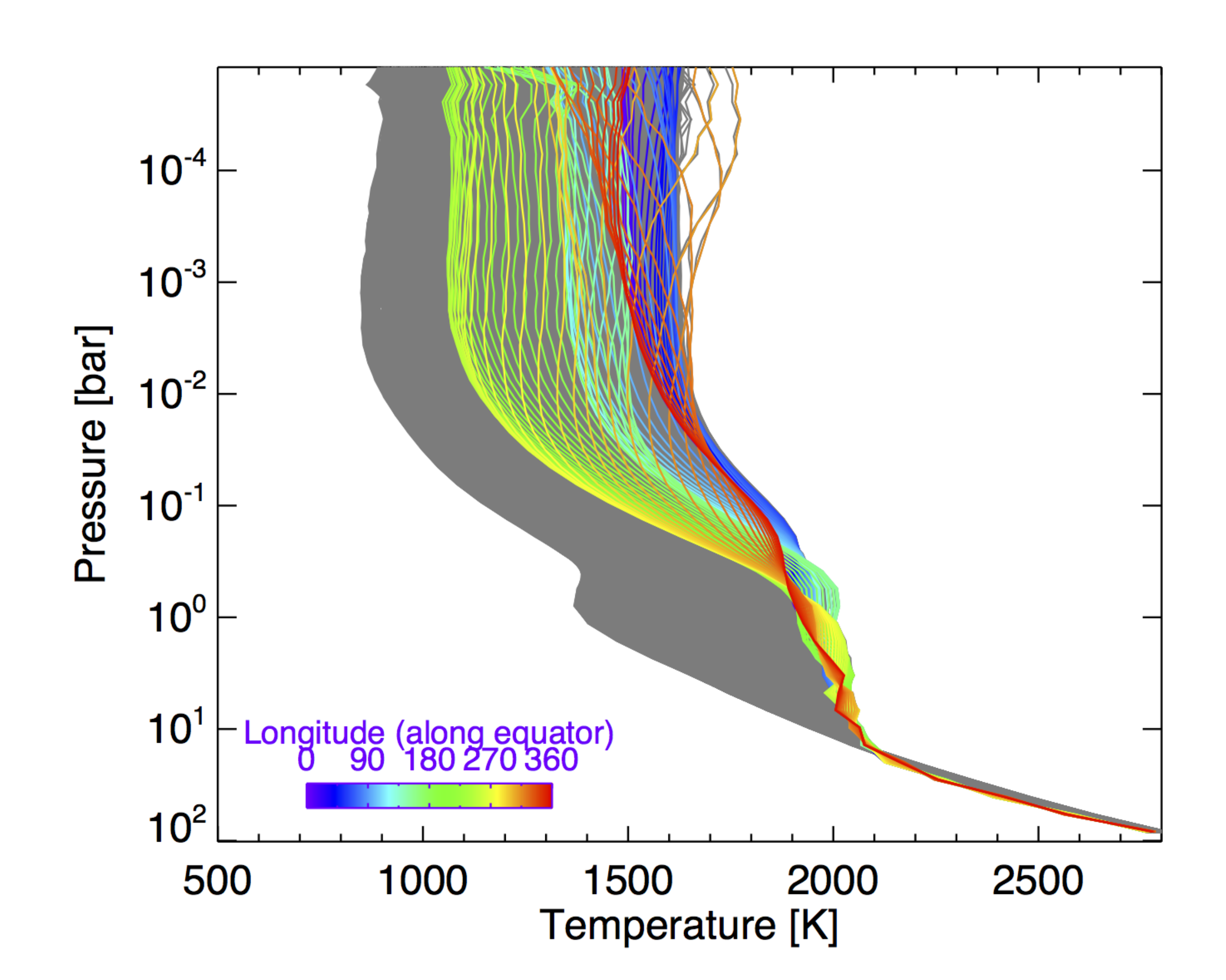}{0.33\textwidth}{HD 209458b}
          \fig{W43_tp.pdf}{0.33\textwidth}{WASP-43b}
          \fig{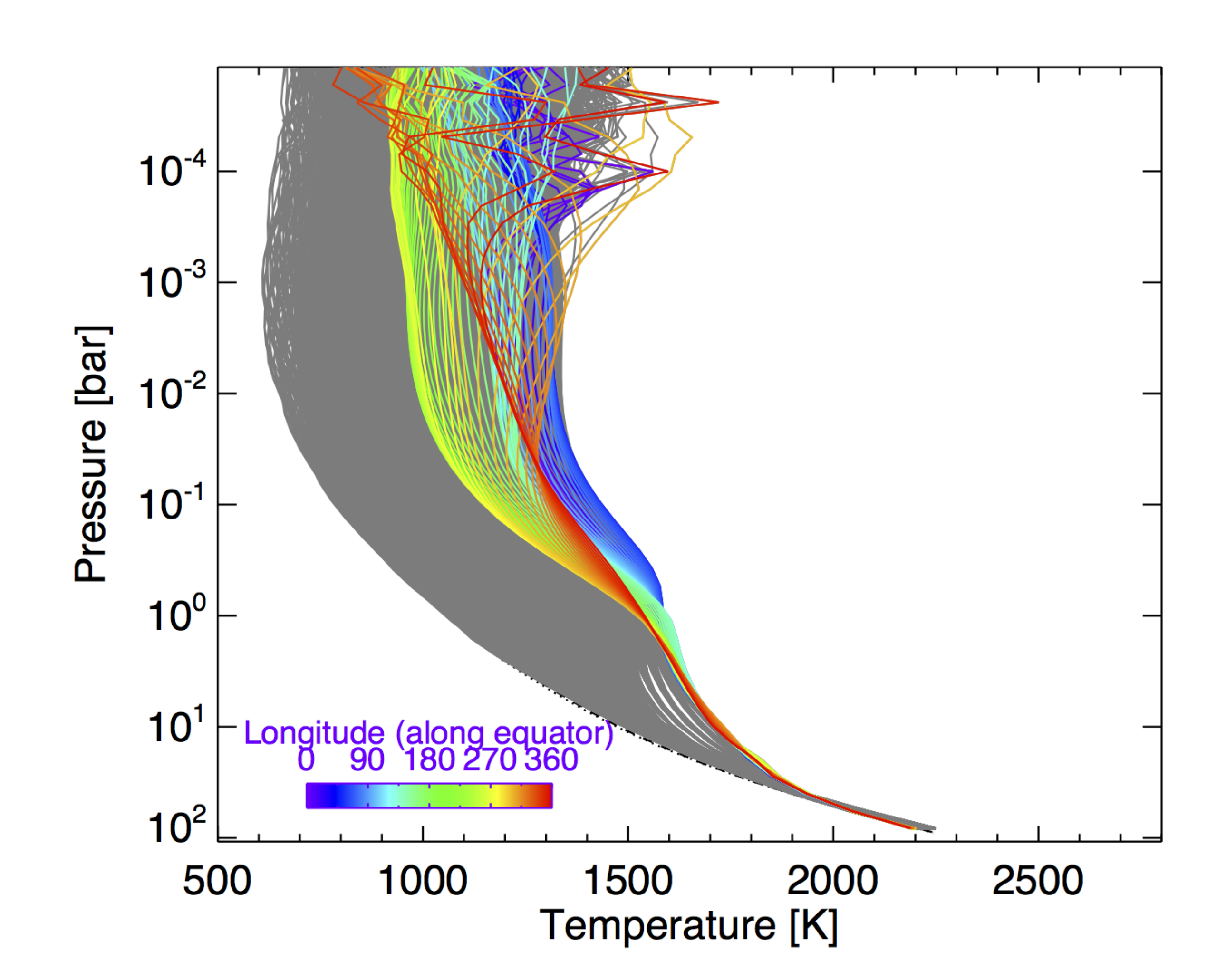}{0.33\textwidth}{HD 189733b}
          }     
\caption{Temperature-pressure profiles drawn at all longitudes and latitudes for three targeted hot Jupiters. Top: vertical temperature-pressure profiles for HD 209458b (purple), WASP-43b (red) and HD 189733b (black). The temperature of WASP-43b spans a wider range compared to HD 189733b and HD 209458b, indicating a larger day-night temperature difference. Bottom: temperature-pressure profiles for each individual hot Jupiter. All non-gray lines are equatorial profiles, taken at longitudes from the substellar point ($\phi=0^\circ$), eastward to the antistellar point ($\phi=180^\circ$), and back to the substellar point as indicated by the line color.
  \label{fig:TP}}
\end{figure*}

Figure \ref{fig:TP} shows the vertical temperature profiles throughout the atmospheres of each planet.  Again we see standard and similar behavior between all three planets: a greater day-night temperature difference higher in the atmosphere and more uniform temperatures at depth, especially around the equator.  HD 189733b has a cooler day side than the other planets, as expected for its lower level of incident stellar flux (see Table 1).  While HD 209458b and WASP-43b have similar incident fluxes and so comparable maximum dayside temperatures, WASP-43b shows a much larger day-night temperature difference than HD 209458b (and HD 189733b). The main contributor to the difference in day-night temperature contrast between WASP-43b and HD 209458b is WASP-43b's enhanced gravity, which is almost six times stronger than on HD 209458b. Roughly speaking, the higher gravity of WASP-43b means that it has a longer Kelvin wave propagation timescale and so transports heat more slowly, while the higher gravity also decreases the radiative timescale of the atmosphere, meaning that the gas can more efficiently cool before being transported away from the substellar point.  The faster rotation rate of WASP-43b (more than four times that of HD 209458b) also contributes to maximizing the day-night temperature contrast through this same comparison of physical timescales, following the analytic analysis of \citet{kom16}.

\subsection{Low Resolution Spectra}

\begin{figure}
\begin{center}
\includegraphics[scale=0.5]{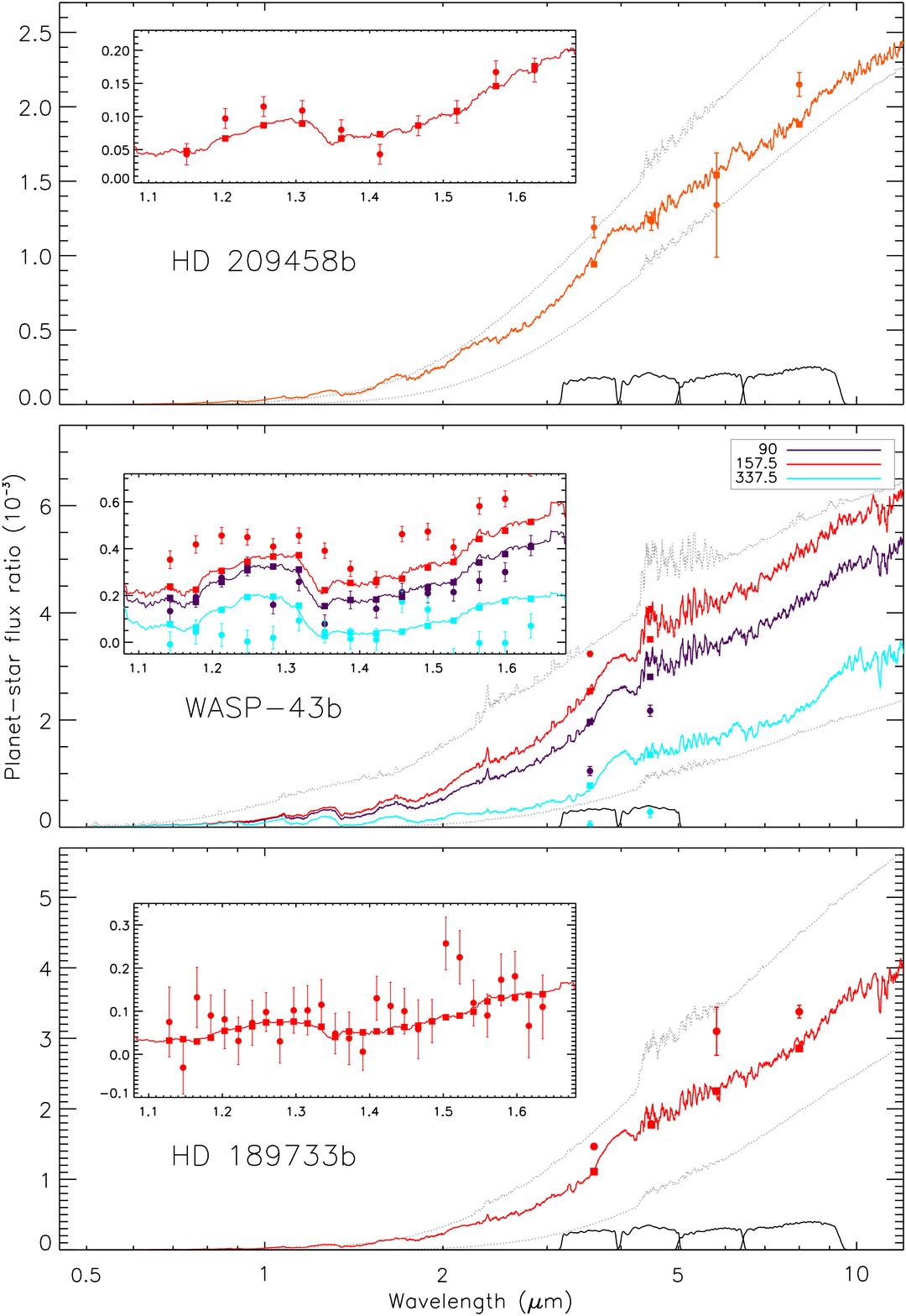}
\end{center}
\caption{Planet-star flux ratio of the three modeled planets at low spectral resolution from 0.5 to 12 microns.  In all three panels, our models are shown with colored lines, isothermal atmosphere predictions (for the planetary spectra) are shown in gray, and observational data from both WFC3 and \textit{Spitzer} IRAC are shown with their reported error bars.  The IRAC bandpasses are shown in black, for reference.   Top: Comparison against observations obtained with WFC3 \citep{line16} and \textit{Spitzer} IRAC \citep{dia14} for HD 209458b. Planetary blackbodies are shown at 1300 K (bottom) and 1600 K (top). Middle: Models and observational data \citep{ste14,ste17} for WASP-43b at orbital phase angles of $90^{\circ}$, $157.5^{\circ}$, and $337.5^{\circ}$. Planetary blackbodies are shown at 1100 K (bottom) and 2200 K (top). Bottom: Dayside emission spectrum of HD 189733b with observations obtained from WFC3 \citep{cro14} and \textit{Spitzer} IRAC (\citet[][]{knu12}, 3.6 and 4.5 $\mu$m; \citet[][]{cha08}, 5.8 $\mu$m; \citet[][]{ago10}, 8.0 $\mu$m). Planetary blackbodies are shown at 900 K (bottom) and 1350 K (top). 
  \label{fig:low_res}}
\end{figure}

We first generate low-resolution emission spectra of the three well-studied hot Jupiters, HD 209458b, WASP-43b and HD 189733b to compare our model results to existing data from \textit{HST} and \textit{Spitzer}. Planet-star flux ratios were calculated using PHOENIX models\footnote{http://phoenix.astro.physik.uni-goettingen.de/} with the best-fit stellar parameters for the three host stars.  Figure~\ref{fig:low_res} shows the comparison to the observational data for all three planets. For both HD 209458b and HD 189733b, the model comparison is made (for the dayside thermal emission spectrum only) to data from \citet{line16}, \citet{dia14}, \citet{cro14}, \citet{cha08}, \citet{knu12}, and \citet{ago10}.  For WASP-43b, a full spectral phase curve is available \citep{ste14,ste17}, and we perform a detailed model comparison at the orbital phases where the authors find the strongest emission ($157.5^{\circ}$), the weakest emission ($337.5^{\circ}$), and at half phase of $90^{\circ}$.  In general, the low-resolution spectra provide information about atmospheric abundances and vertical thermal structures.  The orbital phase curve of WASP-43b also probes the longitudinal temperature structure of the planet (i.e. the location of the hot spot).  However, low spectral resolution does not provide the direct probe of atmospheric dynamics that may be attained at higher spectral resolution.  

In general, our low-resolution spectra for all three planets match the observational data and are consistent with previously published 1-D spectral modeling and GCM results for these objects.  Over the WFC3 wavelength range, our models provide a very good fit to the dayside fluxes of both HD 209458b and HD 189733b, while the \textit{Spitzer} IRAC data provide a more mixed result in terms of matching our models.  For WASP-43b, there is an underestimation of the flux on the dayside and a significant overestimation on the nightside.  This discrepancy was also noted by \citet{kat15}, who found that higher atmospheric metallicity provides a better fit to both the planet's dayside emission spectrum and the observed longitudinal hotspot shift.  (Our own models were only run at solar metallicity.)  The lower than expected night-side flux is postulated by \citet{kat15} to be the result of clouds pushing the night-side photosphere higher in the atmosphere to a cooler location corresponding to the cloud top.  

\subsection{High Resolution Spectra \label{temp}}

\begin{figure}
\begin{center}
\includegraphics[scale=0.32]{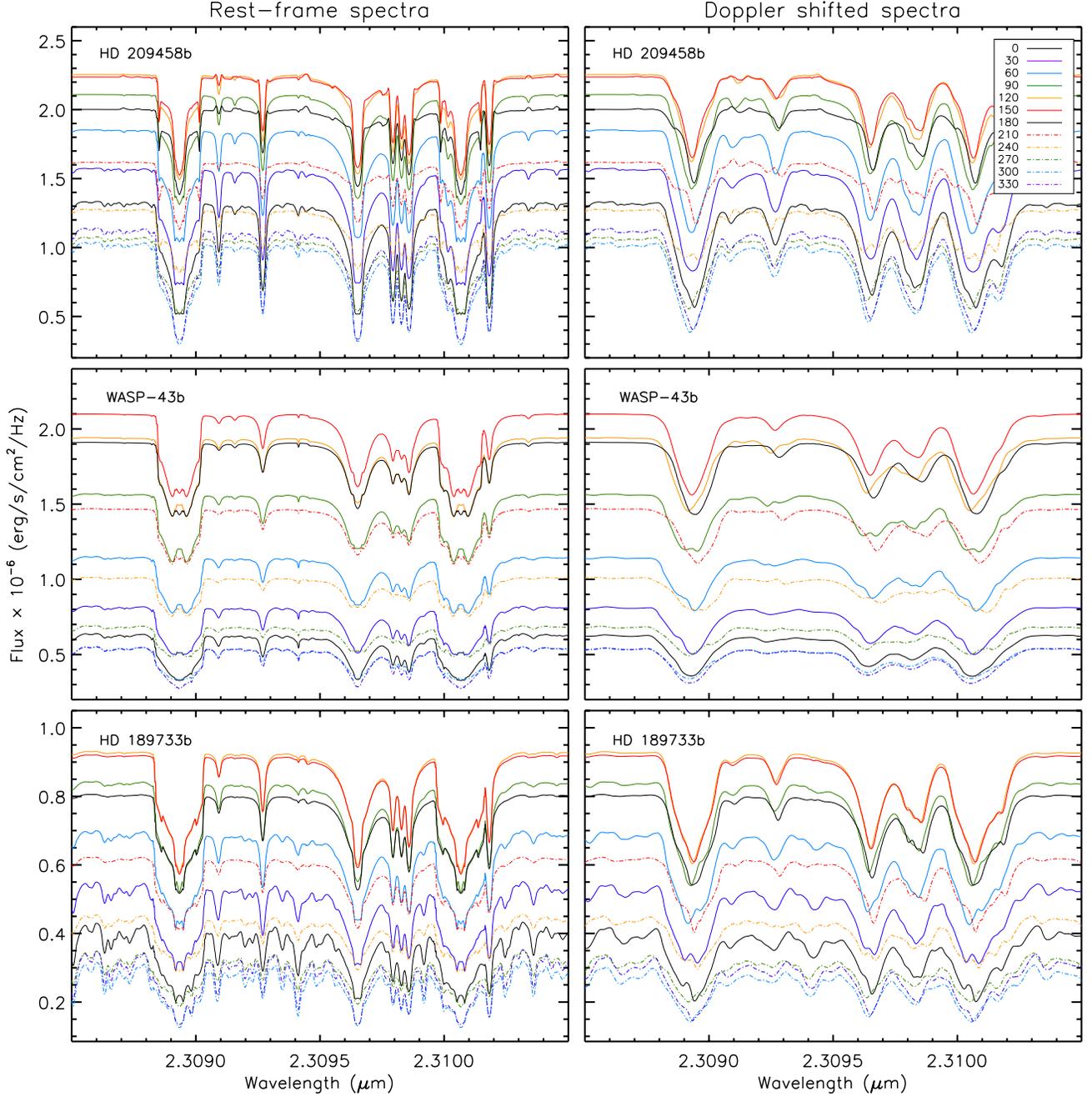}
\end{center}
\caption{Rest-frame and Doppler shifted emission spectra of three targeted hot Jupiters HD 209458b, WASP-43b, and HD 189733b at 12 orbital phase angles from $0^{\circ}$ to $330^{\circ}$. Left panels are rest-frame spectra of the three planets, whereas right panels are Doppler shifted spectra. The continuum intensity of each spectrum is an indicator of temperature. Spectra at phase angles that are equally offset from the substellar point have the same color, clearly indicating that the leading sides of all 3 planets are hotter.
  \label{fig:6_tile}}
\end{figure}

\begin{figure}
\begin{center}
\includegraphics[scale=0.475]{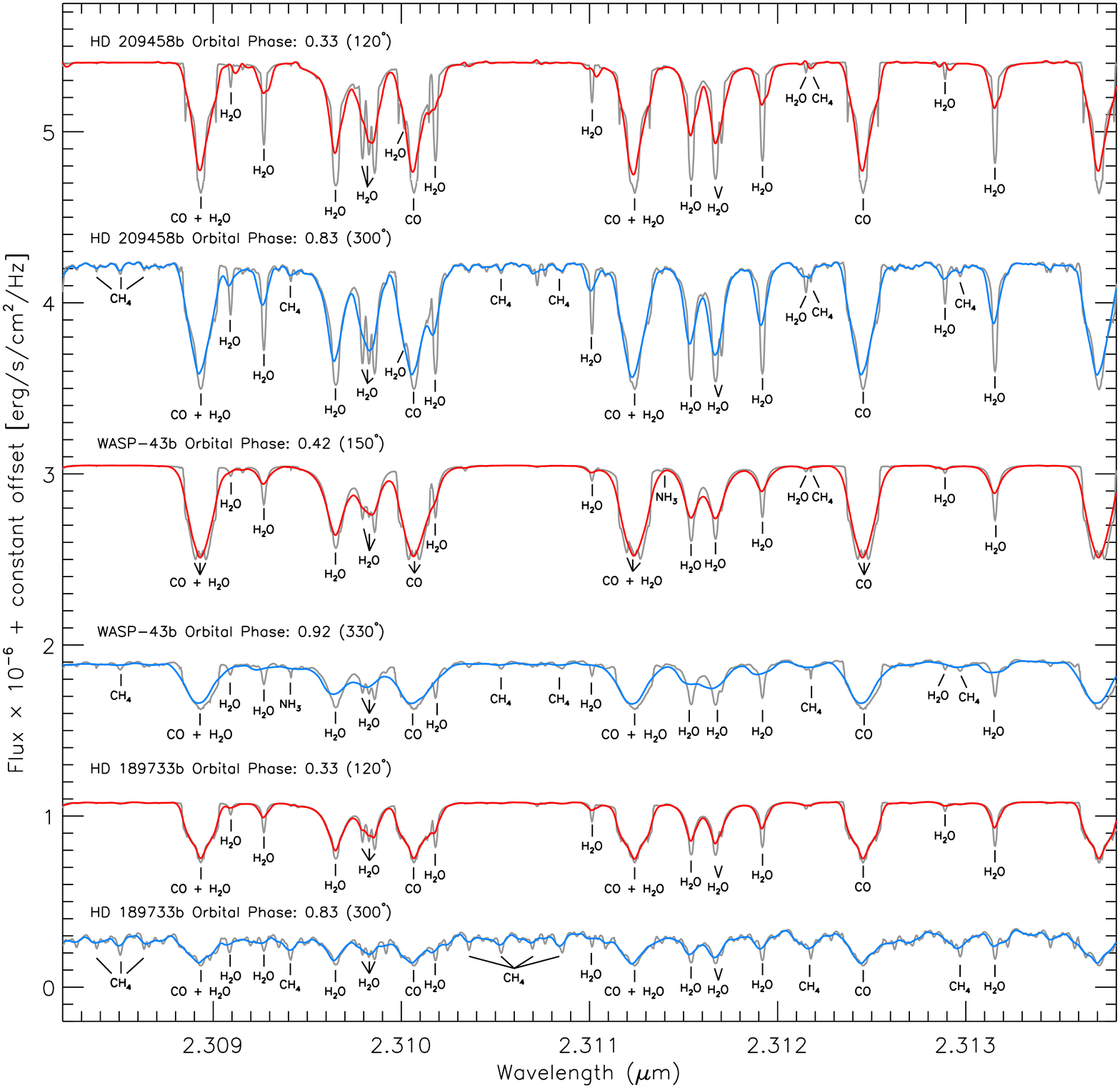}
\end{center}
\caption{Emission spectra for phase angles (as indicated) corresponding to the brightest and dimmest flux for each of the three modeled hot Jupiters. Red lines are Doppler-shifted spectra at the phase angle with the brightest flux. Blue lines are Doppler-shifted spectra corresponding to the dimmest flux. Gray lines are unshifted spectra, shown for reference.  An arbitrary vertical offset is added to each spectrum for easier viewing. Absorption features, which indicate the relative chemical composition and temperature, are labeled. Most unmarked features in the spectra at cooler temperatures are due to methane.   
  \label{fig:offset}}
\end{figure}

A representative section from 2308 to 2314 nm of the high-resolution unshifted (rest frame) and Doppler-shifted emission spectra for all three targeted hot Jupiters are shown in Figure~\ref{fig:6_tile}.  The spectra are generated at 12 equally spaced orbital phase angles to show the phase variations in the spectra.  The full modeled spectral range is shown in Figure~\ref{fig:offset} for each planet at its brightest and dimmest phase angles.   As with the low-resolution spectra, the spectra at high resolution encode information about the atmospheric composition and temperature structure.  In the latter case, the vertical temperature gradient can be detected through the shape of spectral lines (emission vs.~absorption), whereas the longitudinal temperature gradient is measured by the intensity of continuum emission as a function of orbital phase.  HRS has the potential to probe lower pressures than low-resolution spectroscopy because the line cores, formed at high altitude, are fully sampled.  

The 3-D temperature structures for the three planets, as described in Section~\ref{gcm_results}, are revealed through the spectra in Figures~\ref{fig:6_tile} and \ref{fig:offset}.  Both the day-night temperature contrasts and hotspot offsets are discernible from the continuum emission levels.  By comparing the mininum and maximum intensity spectra for each planet in Figure~\ref{fig:6_tile}, the smallest day-night contrast clearly belongs to HD 209458b.  WASP-43b has the largest overall day-night contrast (as seen in Figure~\ref{fig:TP}), although this is mitigated somewhat in the disk-integrated emission spectra because the strong equatorial jet in this planet keeps the low latitude regions quite hot, across all longitudes.  Eastward hotspot offsets are seen in all three planets by noting that the highest level of continuum emission occurs at orbital phase angles less than 180$^{\circ}$, prior to the full dayside rotating into view.  WASP-43b has the smallest hotspot offset of only $\sim$30$^{\circ}$.  The eastward shift of the hotspot can also be seen in Figure~\ref{fig:6_tile} by comparing spectra from phase angles equidistant from 180$^{\circ}$.  In all cases, the smaller phase angle (i.e.~the one prior to the full dayside view) produces substantially higher emission, corresponding to higher disk-integrated brightness temperatures.  Unfortunately, the data reduction process for HRS observations has typically removed any information about the continuum level, so, in that respect, phase curves may remain the best path forward for directly ascertaining hot spot offsets and temperature contrasts in the near future.  In terms of atmospheric composition, HD~189733b, the coolest of the three planets, undergoes a noticeable shift in carbon chemistry from CO on its hot dayside to CH$_4$ on its nightside, whereas portions of the night sides of both HD 209458b and WASP-43b remain hot enough for CO to remain stable.  However, the assumption of local chemical equilibrium could be compromised by the very fast wind speeds on these planets \citep[e.g.][]{coo06}.

\subsection{Doppler Shifts in Emission Spectra}

\subsubsection{Numerical Results \label{results341}}

The Doppler shifted emission spectra generated by our models can be seen in Figures~\ref{fig:6_tile} (right panels) and \ref{fig:offset}.  The Doppler shifted spectra are considerably broadened due to the joint effects of the line-of-sight components of the planet's rotation and wind velocities.  Doppler broadening is especially apparent for WASP-43b due to its fast rotation speed. An additional net blueshift is apparent in the low-intensity spectra in Figure~\ref{fig:offset} (blue lines), which we discuss in more detail below. The comparison between the unshifted and Doppler shifted spectra reveal the need for 3-D modeling at high spectral resolution to accurately replicate the detailed shapes of the spectral lines.

\begin{figure}
\begin{center}
\includegraphics[scale=0.41]{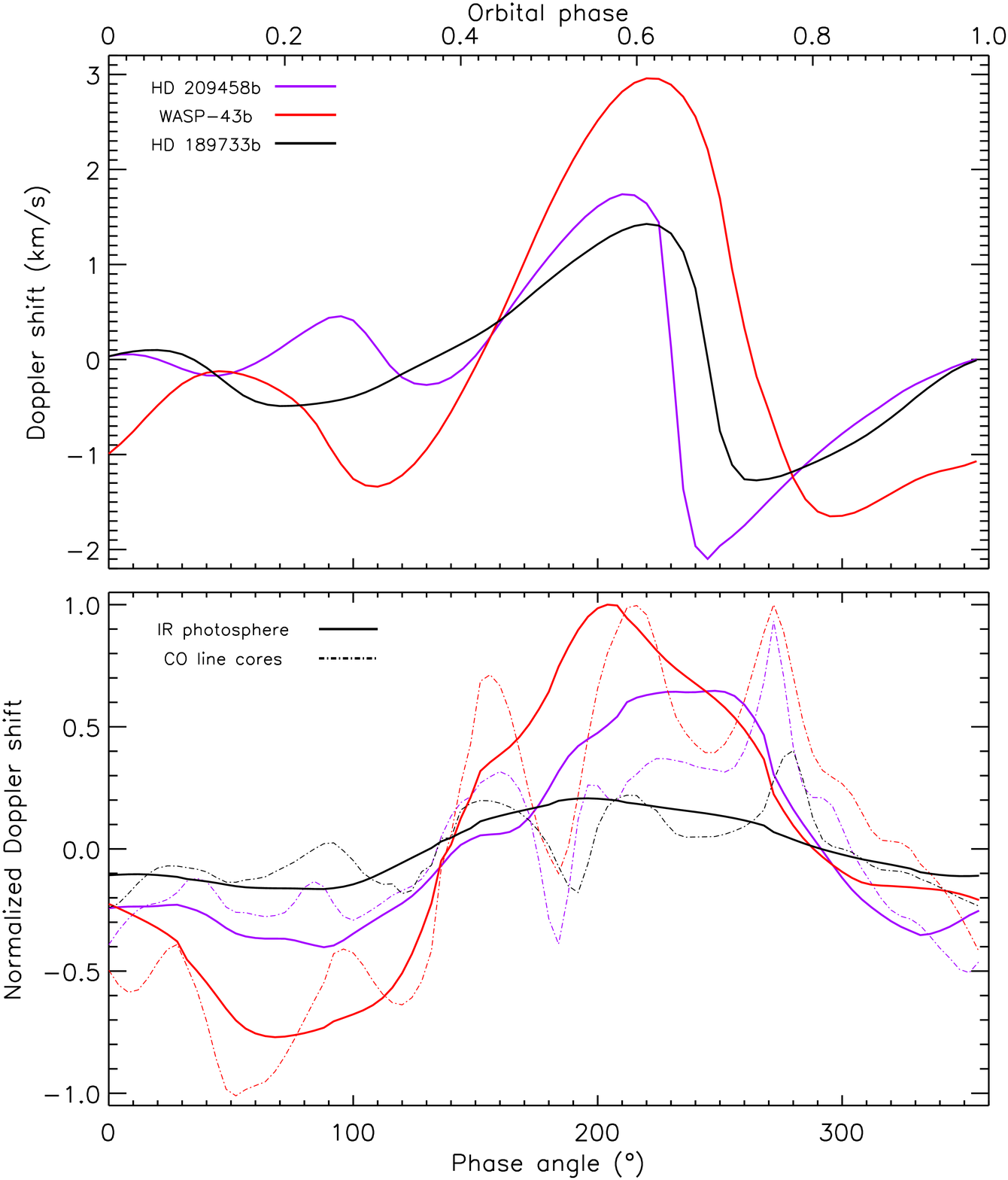}
\end{center}
\caption{Doppler shift of HD 209458b, WASP-43b and HD 189733b as a function of orbital phase. The top panel shows the Doppler shifts obtained by cross-correlating against a rest frame spectrum, and the bottom panel shows the normalized analytic Doppler shifts calculated at pressures corresponding to the IR photosphere (solid line) and CO line cores (dashed line). In general, we obtain a net Doppler blueshift as the hot spot approaches the observer and a redshift as the hot spot recedes. 
\label{fig:DOPPLER}}
\end{figure}

\begin{figure}
\begin{center}

\includegraphics[scale=0.32]{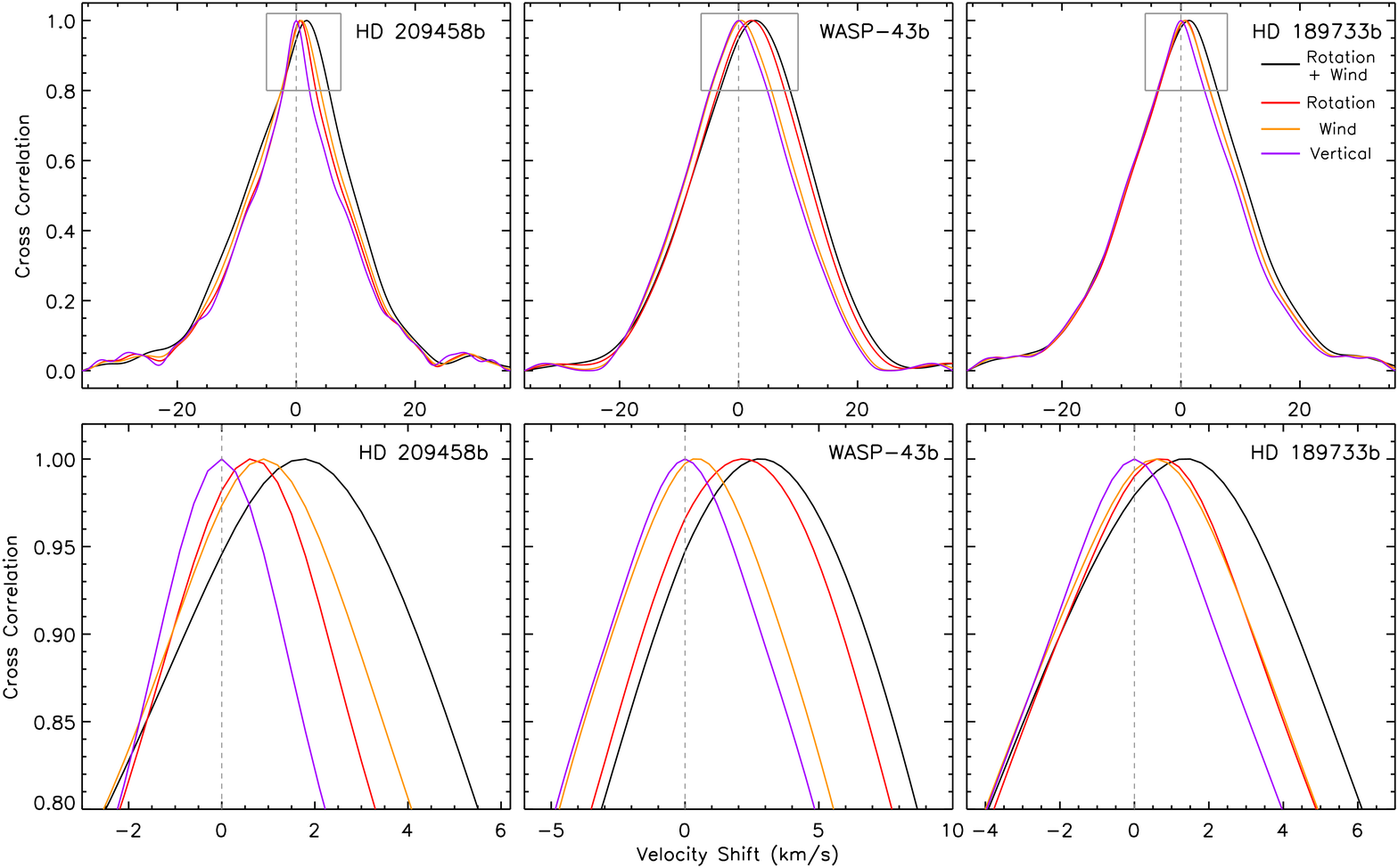}
\end{center}
\caption{Cross-correlation functions for all three planets taken at an orbital phase angle of $210^{\circ}$. The Doppler effects from rotation, line-of-sight winds, and the combination of the two are shown in red, orange, and black, respectively. The effect from the vertical component of line-of-sight wind velocities is shown in purple, which is close to zero for all three planets.
          \label{fig:cross_correlate}}
\end{figure}

To quantitatively assess the magnitude of the Doppler shifts for each model, we have cross correlated our Doppler shifted spectra against their rest frame counterparts, effectively combining the Doppler signatures of each line into a single measurement of net Doppler shift.  The results of this process are shown in Figures~\ref{fig:DOPPLER} (top panel) and \ref{fig:cross_correlate}.  From Figure~\ref{fig:DOPPLER}, it is apparent that non-zero net Doppler shifts are present for each planet, and that the Doppler shift varies substantially as a function of orbital phase.  As seen in Figure~\ref{fig:snapshot}, line of sight velocities originate from a combination of the wind pattern and the planet's rotation.  At the location of the IR photosphere, the strongest redshifted features originate at the receding limb, and the strongest blueshifted features originate at the approaching limb, with a near-symmetric distribution.  The symmetry in the line-of-sight velocity maps could lead to the expectation that Doppler broadening, but no net Doppler shifts, would be present in the disk-integrated emission spectra.  This expectation is incorrect however, due to the asymmetric temperature distribution.  The thermal emission from the hottest location on the planet dominates over the emission from cooler locations on the planet.  The net Doppler shift of the visible side of the planet at a given orbital phase is therefore dominated by the hottest region.  In general, when the planet's hot spot is approaching the observer, a net blueshift is obtained, and when the hot spot is receding, a net redshift is observed.  

The Doppler shifts obtained from the disk-integrated emission spectra are therefore potential probes of both the temperature structure of the planet and the line-of-sight motions that result from both winds and rotation.  However, it is somewhat of a challenge to disentangle the two effects.  In general, the curves shown in Figure~\ref{fig:DOPPLER} show the expected behavior.  At small orbital phase angles, as the night side is in view while the hot spot approaches the observer, net blueshifts are seen in all three planets.  The transition to net redshifts occurs prior to orbital phase 180$^{\circ}$ as the eastward-shifted hot spot begins to move away from the observer.  The maximal redshift occurs near a phase angle of 210$^{\circ}$ when the hot spot becomes strongly confined to the receding side of the planet.  A sharp transition from redshift back to blueshift takes place as the hot spot disappears over the receding limb.  The drop in Doppler shift for HD 209458b and HD 189733b occurs first, followed by WASP-43b. This agrees with predicted trends from both our 3-D models and observations of the three planets \citet{ste17}, \citet{zel14}, \citet{ste14}, and \citet{knu09}.  The strongly peaked redshifted signatures for each planet are a result of the geometry of the equatorial jet, which has a sharper longitudinal gradient in temperature on the receding end than on the approaching end (see Figure~\ref{fig:snapshot}).  Overall, WASP-43b produces the largest net Doppler shifts because it is the fastest rotator of the three planets. 

Smaller-scale features in the Doppler curves in Figure~\ref{fig:DOPPLER} are more difficult to explain qualitatively, but they result from the joint effects of chemistry, atmospheric dynamics, and temperature structure across all altitudes.  The cross correlation signal originates from a comparison of spectral lines, which become optically thick at pressures lower than that of the continuum photosphere.  It is therefore reasonable to assume that the cross correlation function primarily probes atmospheric dynamics at higher altitudes than the band-averaged IR photosphere.  The lower segment of Figure~\ref{fig:snapshot} shows Doppler contours and temperature maps for pressure levels corresponding to the line cores of the strongest CO absorption features (0.14 mbar, 0.85 mbar, and 0.35 mbar, respectively for HD 209458b, WASP-43b, and HD 189733b).  Here we find more asymmetry in the line-of-sight velocities obtained from the approaching and receding hemispheres as well as a somewhat altered temperature pattern where the equatorial jet is less obvious. A comparision between the upper and lower segments of Figure~\ref{fig:snapshot} reveals the complexities in mapping atmospheric dynamics through cross correlation techniques, which will naturally probe a range of pressure levels in the atmosphere.  Additionally, chemical changes may also affect the cross correlation signal, as in the case of the transition in carbon chemistry in HD 189733b where the strong CO lines virtually disappear from the planet's spectrum on its night side.  

To investigate the effects of the individual line-of-sight velocity components on the overall Doppler shift signal, we have separately generated Doppler shifted spectra that include only the vertical or horizontal components of the wind velocities, as well as rotation-only spectra. By cross-correlating these single-component versions of the Doppler shifted emission spectra against the unshifted spectra, we can separately assess their effects.  The cross-correlation functions for all three planets at an orbital phase of $210^{\circ}$, where the Doppler effect is most evident, are shown in Figure~\ref{fig:cross_correlate}. For all three planets, the vertical component of the line-of-sight wind velocities has almost no contribution to the overall Doppler shift, despite providing a direct line-of-sight contribution at the sub-observer point, because the vertical component is approximately two orders of magnitude smaller than the net wind speed. For WASP-43b, the fastest rotator of the three, the major source of the Doppler shift is rotation, with the wind further shifting the emission spectra. For HD 189733b and HD 209458b, the Doppler contributions from wind and rotation are of comparable magnitude.  (Note that in all three cases, the equatorial jet moving in the same direction as the planet's rotation works to reinforce the rotational Doppler shift. Retrograde jets, while not predicted by the GCMs, would have the opposite behavior.) Doppler broadening of the spectra can be assessed by the width of the cross correlation functions.  As expected, the widths of the net cross correlation functions rank with the planets' rotational speeds, with WASP-43b as the broadest and HD 209458b as the narrowest.

We note that our modeling is based on the assumption of circular orbits with an orbital inclination of exactly 90$^{\circ}$. However, all three planets considered in this paper have slightly non-edge-on orbits and marginal non-zero eccentricities, which will induce an additional Doppler shfit in the spectra.  For HD 209458b, WASP-43b, and HD 189733b, the semi-amplitude of the planet's radial velocity will change by 0.217 km/s, 0.055 km/s and 0.056 km/s, respectively, given their non-zero eccentricities.  Incorrect subtraction of the orbital Doppler shift from the planet's emission spectrum at the $\sim$$0.1 - 1$ km/s level could lead to a misinterpretation of the planet's Doppler shift signature due to winds and rotation.  This could occur, for example, if the orbital eccentricity of the planet in question is not very well constrained. Additionally, if the spurious radial velocity signal due to an eccentric orbit reaches its maximum or minimum around the same phases as the maximum and minumum predicted Doppler shifts due to winds and planetary rotation, it would be possible to confuse the two effects. A well-constrained measurement of planet's angle of periastron from radial velocity data would be required to disentangle the effects.

%We carried out the calculation of semi-amplitute radial velocities for these planets using known eccentricity\footnote{http://exoplanet.eu} and computed resulted extra velocity shifts. Extra velocity shifts are 0.217 km/s, 0.055 km/s and 0.056 km/s for HD 209458b, WASP-43b and HD 189733b respectively. In other words, errors in orbital velocities for WASP-43b and HD 189733b should not have an distinctive effect on Doppler shifts at different orbital phase angles, whereas that for HD 209458b could lead to a more questionable result. In addition, if the spurious radial velocity signal due to an eccentric orbit reaches its maximum and minimum around the same phases as the maximum and minumum predicted Doppler shifts due to winds and planetary rotations, it would be possible to confuse two effects. A measurement of planet's angle of periastron, which should be well-constrained by stellar radial velocity data, would be required to disentangle the effects. }

\subsubsection{Analytic Estimates}

As we mentioned before, information from net Doppler shifts of emission spectra on wind speed, rotation rate and temperature is entangled due to the emission spectra's dependency on the thermal structure of the exoplanet.  
However, we can estimate the strength of the net Doppler shift due to each component and use this to interpret the predicted signal we report in the top panel of Figure~\ref{fig:DOPPLER}.  If we assume that the Doppler shift from each region of a planet's atmosphere goes as $D_{net}\sim F \times v_{los}$, then we can combine the geometric factors containing latitude and longitude in the expression for the line-of-sight velocity (Equation~\ref{eqn:vlos}) with the geometric factors used to weight each region of the planet when integrating over the observed disk in order to identify which regions of the planet contribute the most to the net Doppler signal.  If we assume that the east-west winds dominate over the north-south and vertical ones (largely valid), then the equatorial regions at $\pm45\degr$ of the sub-observer longitude will give us the largest signal, just due to geometry.  This assumes otherwise constant wind speeds and temperatures across the planet, which is obviously not the case, but gives us a way to simplify the situation such that we can express the expected net Doppler shift as:
\begin{equation}
D_{net} \sim R_p \Omega [F(\phi_{\mathrm{obs}}+45^\circ) - F(\phi_{\mathrm{obs}}-45^\circ)]                              + u(\phi_{\mathrm{obs}}+45^\circ) F(\phi_{\mathrm{obs}}+45^\circ)                                -  u(\phi_{\mathrm{obs}}-45^\circ) F(\phi_{\mathrm{obs}}-45^\circ) \label{eqn:simpleDnet}
\end{equation}
where the flux and wind velocities are implicitly equatorial values. For synchronous rotation the sub-observer longitude is directly related to the orbital phase angle by: $\phi_{\mathrm{obs}}=180^\circ-\varphi$.

We can then use this expression, with the further assumption of blackbody emission (so $F \rightarrow T^4$), to calculate the expected net Doppler shift from some particular layer of the atmosphere, by using the temperature and wind field at that pressure level.  In the bottom panel of Figure~\ref{fig:DOPPLER} we plot the simplified expected Doppler shift, using Equation \ref{eqn:simpleDnet} for each planet at its infrared photosphere (i.e. the pressure from which most of the emission is produced) and at the pressure level corresponding to the CO line cores (i.e. where the strongest spectral lines are produced).  These are also the pressure levels shown in Figure~\ref{fig:snapshot}.  While this simplistic expression is unsurprisingly not able to fully reproduce the results of the much more detailed calculation, we do recover some of the more general features.  The simplified curves for the photospheric levels are much smoother than those for the CO line cores (the result of the smaller scale features seen in the wind and temperature patterns at the lower pressure levels), while the actual signal is somewhat intermediate between the two.  This is in line with our expectation that the actual signal observed will be a detailed combination of emission across multiple pressure layers, with a complex weighting for the net Doppler shift.

One advantage of this simplistic expression, which can roughly reproduce the correct answer, is that it gives us a framework in which we could combine high-resolution and broadband thermal phase curves for a planet in order to retrieve a rough profile of the planet's wind speed as a function of longitude.  For a transiting planet we know its radius and we can assume synchronous rotation for hot Jupiters in order to calculate its rotation rate.  We could then use Equation~\ref{eqn:simpleDnet} to combine the net Doppler shift as a function of orbital phase angle (from high-resolution orbital spectroscopy) and the thermal flux as a function of observed longitude (from a broadband thermal phase curve) to solve for east-west wind speed as a function of longitude.  Those combined data sets perhaps contain enough data to drop the assumption of synchronous rotation and solve for $\Omega$ as well (accounting for the now more complex relation between flux emitted from some longitude and flux observed as a function of time).  The retrieved wind profile would be imprecise, due to the simplifying assumptions as well as any potential mismatch between the pressure layers probed by the different observations, but this would nevertheless provide a unique constraint on atmospheric circulation models.

\section{Conclusions \label{conclusion}}

We have developed a coupled atmospheric dynamics model and emission spectroscopy radiative transfer code to study the effects of winds and planetary rotation on the high-resolution emission spectra of the three benchmark hot Jupiters, HD 209458b, WASP-43b and HD 189733b. This is the first time that an emission spectrum radiative transfer model based on 3-D GCM output has self-consistently treated the line-of-sight geometry, including the effects of global wind patterns and rotation. We  have demonstrated how chemistry, thermal structures, winds, and rotation can be inferred through disk-integrated emission spectra and their associated Doppler shift signatures. Our models compare favorably with previously published observations and models of the thermal emission from the three targeted planets.  

While exoplanet atmospheric dynamics have traditionally been constrained through IR phase curve observations aimed at measuring hotspot offsets and day-night temperature contrasts, Doppler shifted emission spectra provide a direct probe of wind speeds and rotation rates.  We find that all three modeled hot Jupiters share similar trends in net Doppler shifts as a function of orbital phase, with magnitudes of up to several km~s$^{-1}$. The Doppler shift signatures are tied to the thermal structure of the equatorial jet, with major peaks and troughs indicating the appearance and disappearance of hotter regions.  While exact predictions of Doppler shifts in disk-averaged emission spectra require full 3-D atmospheric dynamics models like the ones presented in this paper, we have laid out a simple method for estimating equatorial wind speeds by combined measurements of Doppler shifted emission spectra and IR phase curves.  It is also necessary to point out that the absence of Doppler shifts in 1-D models could lead to radial velocity offsets and systematic biases in interpreting HRS observations.  

High-resolution spectroscopy (HRS) is a relative newcomer to the list of observational techniques being used to successfully characterize exoplanet atmospheres.  The appeal of this method from the ground is that telluric contamination can be removed in a fairly straightforward manner, and the technique can provide complementary information to high-precision but lower spectral resolution data obtained from space-based facilities \citep[see e.g.][]{bro17}.  The CRIRES spectrograph \citep{kae04} on the VLT has been the forefront instrument for HRS of exoplanet atmospheres.  With a resolution of up to 10$^{5}$, the instrument can provide Doppler shift information on the km~s$^{-1}$ level.  CRIRES was taken offline in 2014 but will be replaced with the refurbished CRIRES$+$ instrument in 2018.  The upgraded instrument will provide up to a factor of 10 in increased wavelength coverage accompanied by equivalent or better throughput than the first-generation version of the spectrograph, which will allow for atmospheric characterization of planets orbiting fainter target stars.  Other high-resolution spectrographs with the potential to characterize hot Jupiter atmospheric dynamics that are currently under development or in commissioning include SPIRou ($R = 70,000$), iShell ($R = 70,000$), and CARMENES ($R = 80,000$), albeit all three instruments are associated with smaller 3-4 meter telescopes.  In the era of 30-meter class telescopes, the GMTNIRS instrument on the Giant Magellan Telescope will provide $R \sim 100,000$ spectra, opening the door to ground-based atmospheric characterization of smaller and fainter systems.

Finally, we remind the reader that our model has so far explored three transiting hot Jupiters under the assumption that they are tidally locked. However, the tidal dissipation process for gas giants is complex, and no observations have yet fully confirmed the assumption of synchronized rotation. At high-resolution, emission spectroscopy has the potential to provide a direct estimation of the rotation rate, accompanied by information on detailed wind patterns and thermal structures. Further investigation of the Doppler shift signatures in thermal emission spectra of non-synchronously rotating hot Jupiters is therefore warranted. Moreover, the power of high-resolution emission spectroscopy goes far beyond transiting exoplanets. The technique is already proven for measuring the true masses, inclinations, and chemical compositions of non-transiting exoplanets. Modeling of these planets at high spectral resolution is needed for proper interpretation and will become increasingly important as more and better data become available from soon-to-be commissioned HRS instruments. 

\acknowledgements 
We thank an anonymous referee and Matteo Brogi for helpful comments on this manuscript.  This research was supported by NASA Astrophysics Theory Program grant NNX17AG25G.  JZ was additionally supported by the Grinnell College Mentored Advanced Project (MAP) program.  EMRK received support from the Grinnell College Harris Faculty Fellowship.

\bibliography{ms}

\end{document}